\documentclass[aps,prl,notitlepage,twocolumn,superscriptaddress,nolongbibliography]{revtex4-2}
\usepackage{amsmath}
\usepackage{amssymb}
\usepackage{bbold}
\usepackage{color}
\usepackage{tikz}
\usepackage{pgfplots}
\pgfplotsset{compat=1.5}
\usepackage{graphicx}
\usepackage[colorlinks=true, linkcolor=blue, citecolor=blue, pdfencoding=auto]{hyperref}
\usepackage{blindtext}
\usepackage{physics}

\def\Q{\mathcal{Q}}

\def\inv{^{-1}}

\def\phdag{{\phantom \dagger}}

\begin{document}
\title{
Quantum Metric in Step Response
}

\author{Nishchhal \surname{Verma} }
\affiliation{Department of Physics, Columbia University, New York, NY 10027, USA}

\author{Raquel \surname{Queiroz} }
\email{raquel.queiroz@columbia.edu}
\affiliation{Department of Physics, Columbia University, New York, NY 10027, USA}
\affiliation{Center for Computational Quantum Physics, Flatiron Institute, New York, New York 10010, USA}

\begin{abstract}
Quantum geometry of Bloch wavefunctions has gained considerable interest with the discovery of moiré materials that exhibit bands flattened by quantum interference. The quantum metric, the symmetric part of the quantum geometric tensor, influences several observables, such as the dielectric constant, superfluid stiffness and optical spectral weight. However, a direct measurement of the metric itself has remained elusive so far.
In linear response functions such as the conductivity, the matrix elements of the metric typically appear convoluted with energy prefactors, preventing finding an observable that is directly proportional to the total quantum metric.
The only observable that may extract it is the integrated optical spectral weight weighted by the inverse frequency, a generalized sum rule known as the Souza-Wilkens-Martin (SWM) sum rule. However, the sum rule comes with experimental challenges, such as requiring a large spectrum of frequency resolution.
In this work, we propose relaxation from constrained equilibrium as a method to directly measure the symmetric part of the time-dependent quantum geometric tensor (tQGT), which at $t=0$ is the quantum metric. Additionally, we comment on other geometric properties of insulators that are absent in the frequency expansions of conductivity in insulators but can, in principle, be revealed in step response.
\end{abstract}

\maketitle

\paragraph*{Introduction.---}
To first approximation, insulators are materials in which electrons are tightly bound to the ions. These electrons do not form conducting channels and only exhibit small fluctuations around the atom, leading to trivial temporal dynamics.
Topological insulators challenge this notion \cite{Hasan2010}. The electrons exhibit fluctuations at the scale of the unit cell and yet lead to robust edge phenomena with quantized transport, similar to Landau levels.
Topology obstructs the local picture of bound electrons in a topological insulator \cite{Monaco2018}.

We can understand the zero-point motion of bound electrons as the generator of quantum geometry in the electron wavefunction \cite{verma2024instantaneous,onishi2024quantum}. 
Unlike topology, quantum geometry is ubiquitous in materials and affects many ground state properties and excitations. For instance, it affects the mass of bound states \cite{Julku2016,Torma2021}, electron-phonon coupling \cite{yu2023}, and may even play an important role in the search for high-temperature superconductors \cite{Aoki2020}. 
The discovery of moiré materials has further brought quantum geometry into the spotlight \cite{Andrei2021}. The flat bands emerge from the destructive interference of electronic paths over the moiré length scale, leading to substantial zero-point motion and, hence, large quantum geometry.
While indirect effects of quantum geometry are ubiquitous \cite{Abouelkomsan2023,kwan2021}, it is worth noting that the metric itself has never been directly measured.

This work aims to find an experimental setup that can directly measure the symmetric part of the quantum geometric tensor, that is, the quantum metric. A hint towards the setup comes from the Souza-Wilkens-Martin sum rule \cite{Souza2000} where one can identify $\sigma(\omega)/\omega$ as a response to an electric field that follows a $1/\omega$ frequency dependence. Alternatively, in the time domain, it corresponds to a constant electric field which abruptly goes to zero at $t=0$, that is $E(t)\propto\Theta(-t)$.
In the following, we will show that relaxation from constrained equilibrium provides a setting to obtain quantum metric. Interestingly, step response also includes geometric quantities that are otherwise absent in conductivity, such as the orbital magnetic moment\cite{Thonhauser2005,Souza2008}.

We begin by reviewing the quantum geometric tensor (QGT) and quantum metric in linear response theory. We then introduce the time-dependent QGT (tQGT) as a unifying principle for the geometric properties of insulators via sum rules.
Next, we discuss linear response theory and the concept of step response. We derive a generalized fluctuation-dissipation theorem (FDT) for the virtual contribution to the dipole-dipole correlation. This derivation crucially depends on the fermionic nature of the distribution function.
The generalized FDT relates symmetric and antisymmetric response functions at high temperatures, providing a framework for studying both symmetric and antisymmetric components of the tQGT.

\paragraph*{Quantum metric in Linear Response.---}
As we discussed, the zero-point motion of electrons in insulators traces its origin back to the fundamental relation between position and momentum in quantum mechanics.
It is well known that electric fields couple to the dipole operator and lead to dipolar fluctuations in isolated atoms \cite{carmichael1999two}.
A generalization to electrons in a lattice involves the virtual dipole-dipole correlation function
\begin{equation}
    \mathcal{Q}_{\mu\nu}(t-t') = \left\langle \hat{r}_\mu (t) \; \hat{Q} \; \hat{r}_\nu(t') \right\rangle \label{eq:def-Q-mu-nu-t}
\end{equation}
where $\hat{r}$ is the position operator and $\hat{Q}$ is the projector into unoccupied states, and therefore its inclusion in \eqref{eq:def-Q-mu-nu-t} selects the \emph{inter-band} dipole transitions. 
The expectation value is taken in the ground state $\langle \cdot \rangle = {\rm Tr}[\hat{P} \cdot]$ where $\hat{P} = 1- \hat{Q}$ is the ground state projector. 
For Bloch electrons, the $t=0$ value corresponds to the quantum geometric tensor with quantum metric $g_{\mu\nu} = {\rm Re}[\mathcal{Q}_{\mu\nu}(0)]$ and Berry curvature $\Omega_{\mu\nu} = {\rm Im}[\mathcal{Q}_{\mu\nu}(0)]/2$.
The tensor in eq.~\eqref{eq:def-Q-mu-nu-t} is therefore named the time-dependent quantum geometric tensor (tQGT)\cite{komissarov2024quantum,verma2024instantaneous}.

Introducing time dependence in Eq.~\eqref{eq:def-Q-mu-nu-t} serves to unify several geometric properties of materials in one formalism. Namely, it was shown that the orbital magnetic moment \cite{Thonhauser2005}, dielectric function \cite{komissarov2024quantum}, and optical mass \cite{Verma2021,Mao2023,Kruchkov2023} appear as various time derivatives of Eq.~\eqref{eq:def-Q-mu-nu-t} and can be effectively extracted via generalized sum rules of optical conductivity $\sigma(\omega)$ \cite{verma2024instantaneous}.
The reason behind this unification is that the anti-symmetric part of the tQGT $\mathcal{Q}^{\rm as}_{\mu\nu}(t)\!=\! \mathcal{Q}_{\mu\nu}(t)\!-\!\mathcal{Q}_{\mu\nu}(t)^\dag$ can be extracted from linear response
and it is related to the optical conductivity by
\begin{equation}
    \sigma_{\mu\nu}(t) = \dfrac{\pi e^2}{\hbar} \; \Theta(t) \; \partial_t \mathcal{Q}^{\rm as}_{\mu\nu}(t). \label{eq:sigma-Q-relation-t}
\end{equation}
The formalism outlines a consistent way to get relations between various sum rules, which have proved useful for a variety of materials \cite{komissarov2024quantum,onishi2023quantum,onishi2024universal, ghosh2024probing}. 
Crucially, all geometric properties of insulators are not independent of each other and are fixed by the definition of the projectors $\hat{P}$ and $\hat{Q}$. This realization also permits a consistent interpolation between \emph{ab-initio} and tight-binding methods, by a controlled truncation of the Hilbert space. 

Eq.~\eqref{eq:sigma-Q-relation-t} suggests that the Kubo formula for conductivity contains the matrix elements for quantum geometry. 
It can be seen explicitly in the Fourier domain when we expand in powers of frequency $\omega$,
\begin{equation}
\sigma_{\mu \nu}(\omega) = \frac{e^2}{i\hbar} \sum\limits_{m \neq n} f_{n m} \omega_{m n} \hat{r}^{nm}_\mu \hat{r}^{mn}_\nu \sum\limits_{p=0}^\infty\left[\dfrac{\omega}{\omega_{mn}}\right]^p \label{eq:cond-omega-Taylor}
\end{equation}
where $f_{nm}=f_n-f_m$ and $f_n$ are the occupation factors, $\hbar\omega_{mn}=E_n-E_m$ are the energy differences between states and $\hat{r}_{nm}^\mu \hat{r}_{mn}^\nu$ are the matrix elements of the position operator.
The power series in $\omega/\omega_{mn}$ arises from the Taylor expansion of $1/(\omega-\omega_{mn})$.
The position matrix elements can be split into symmetric and anti-symmetric combinations $\hat{r}^{nm}_\mu \hat{r}^{mn}_\nu = g^{m n}_{\mu \nu} + i \Omega^{m n}_{\mu \nu}/2$ where $g^{m n}_{\mu \nu}$ and $\Omega^{m n}_{\mu \nu}$ are related to the quantum metric and Berry curvature. It is instructive to notice that the anti-symmetry of $f_{nm}$ and $\omega_{mn}$ forces the metric $g^{m n}_{\mu \nu}$ contribution to vanish at zeroth order in frequency.
More generally, the metric matrix elements always appear in the conductivity with odd powers of $\omega_{mn}$ and, therefore, they always appear convoluted by energy prefactors in nonresonant response.
This fact makes it a challenge to find the total quantum metric $g_{\mu\nu}=\sum_{m,n} f_n(1-f_m) g_{\mu\nu}^{mn}$ directly in a linear response function. 
Similarly, in non-linear response, matrix elements of $g_{mn}^{\mu\nu}$ appear naturally in resonant responses such as in shift and injection currents \cite{Ahn2022}, but never in an integrated form in a non-resonant response.

An exception can be made for Landau levels of free electrons. Here, the only nonvanishing dipole matrix elements are between consecutive levels $\hat{r}^{mn} \propto \delta_{m,n\pm 1}$ leading to a single transition frequency $\omega_c = eB/m$, and therefore $\omega_{mn}$ in Eq.~\eqref{eq:cond-omega-Taylor} can be factored out of the sum. This results in non-resonant optical conductivity that captures both the integrated quantum metric and Berry curvature to all orders.

Correlations can also help factor out the energy prefactor.
Superfluid stiffness, which appears as the weight of the delta function in optical conductivity, is given by the quantum metric times an interaction scale in a flat band superconductor within mean-field theory \cite{Torma2021}. While the result is sensitive to competition with other correlated states \cite{Hofmann2020} and applies for exactly flat bands in certain lattices, it raises several important questions about geometric contributions to the effective mass of the Cooper pairs \cite{Torma2018,Iskin2018,Iskin2021,Verma2021}.
Notably, it presents a clear violation of the Ferell-Glover-Tinkham sum rule \cite{Ferrell1958,Tinkham1959}.
Similar analysis has also been done for excitons, which are electron-hole bound states \cite{Hu2022,Hu2023,Verma2024exciton}.
Other than correlations, it has been suggested that disorder can extract quantum metric in DC response. 
In the limit where electron-scattering $\tau\inv$ rate is larger than the bandwidth $w$, the DC conductivity was shown to carry a universal quantum metric contribution \cite{Mitscherling2022}. 
It was later pointed out that the order of limits $w \rightarrow 0$ and $\tau \rightarrow 0$ is crucial in obtaining the said universal contribution \cite{Huhtinen2023}.
Overall, there has been mounting evidence that quantum metric is hidden in matrix elements of the optical conductivity, and it is revealed only when an interaction or disorder scales the energy factors out of the Kubo formula. 

\begin{figure}
    \centering
    \includegraphics[width=0.49\textwidth]{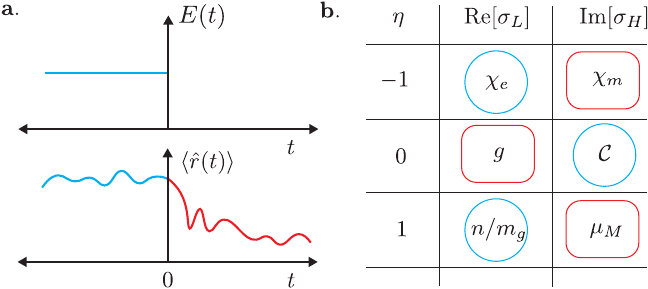}
    \caption{
    {\bf a}.~Schematic showing the step response setup with a drive $f(t)$ that is slowly turned on from $-\infty$ and is shut down at $t=0$. The system then relaxes from a constrained equilibrium state to the ground state. {\bf b}.~Quantum geometric properties obtained from various generalized sum rules of real and imaginary parts of longitudinal and Hall conductivity with various $\eta$. The circled quantities appear in frequency expansions as a result of Kramers-Kronig relations, as opposed to the boxed quantities.
    }
    \label{fig1:schematic}
\end{figure}

The energy scale can also be introduced in sum rules with frequency pre-factors.
This is exactly what the Souza-Wilkens-Martin sum rule accomplishes: the integral over all positive frequencies of the optical conductivity weighted by $1/\omega$ is exactly equal to the quantum metric, independent of the dispersion.
Crucially, the SWM sum rule includes only the positive frequencies in its integral. If the integral were taken over all frequencies, the sum rule would be amenable to Kramers-Kronig relations and make an appearance in the imaginary part of the conductivity. The restriction to positive frequencies does not allow that to happen.
This feature is not limited to quantum metric but all quantities shown marked in red in Fig.~\ref{fig1:schematic}{\bf b}. They are accessible only via generalized sum rules but not in low-frequency expansions of linear response \cite{verma2024instantaneous}.

\paragraph*{tQGT in Insulators.---}
Gapless systems like metals have a diverging quantum metric owing not to interband transition but rather to the presence of a Fermi surface \cite{Resta2006,Resta2011}. Therefore, this work focuses on insulators with a finite gap.
The gap removes intraband diagonal matrix elements of the position operator (which have a gauge redundancy), reducing the expression to \emph{inter-band} terms only.
While the formalism is quite general and applies to any gapped quantum system \cite{verma2024instantaneous}, we will consider non-interacting electrons for a simpler presentation. It means that the system admits a Bloch represetation with states $|\psi_{m, {\bf k}}\rangle = e^{i {\bf k}\cdot \hat{ {\bf r} } } |u_{m,{\bf k} } \rangle$ where $m$ is the band and ${\bf k}$ is the crystal momentum.
Expanding Eq.~\eqref{eq:def-Q-mu-nu-t} in the cell periodic part of the Bloch states, we get
\begin{equation}
    \mathcal{Q}_{\mu\nu}(t) = \int_{\bf k} \sum\limits_{m\neq n} f_{n} (1-f_{m}) \hat{r}_\mu^{nm} \hat{r}_\nu^{mn} e^{ i \hbar \omega_{mn} t }
\end{equation}
where $\hat{r}_\mu^{nm} = \langle u_{n, {\bf k} } | \hat{r} | u_{m, {\bf k} } \rangle $ are the position matrix elements given by Blount \cite{Blount1962}, $\hbar \omega_{mn} = E_{m, {\bf k}} - E_{n, {\bf k}}$ are the band energies. 
We have suppressed momentum labels for brevity, but it should be noted that all terms in the equation have the same momentum ${\bf k}$.
The occupation function $f_n$ is the usual Fermi factor $f[E] = 1/(1+e^{\beta (E-\mu)})$.

Many geometric properties of insulators arise from the time derivatives of the tQGT,
\begin{equation}
   \mathcal{S}_{\mu\nu}^\eta =\int\limits_0^\infty d\omega \; \dfrac{ \sigma^{\rm abs}_{\mu\nu}(\omega) }{ \omega^{1-\eta} } = \dfrac{\pi e^2}{\hbar} \left[ (-i \hat{\partial}_t)^\eta \mathcal{Q}_{\mu\nu}(t) \right]_{t=0} \label{eq:old-main-result}
\end{equation}
with the absorptive part of the conductivity being $\sigma^{\rm abs}_{\mu\nu} = (\sigma_{\mu\nu} + \sigma_{\nu\mu}^*)/2$ \cite{verma2024instantaneous}. 
Sum rules for different $\eta$ have longitudinal and Hall parts, all of which are summarized in Fig.~\ref{fig1:schematic}{\bf b}.
Most relevant are the $\eta=0$ sum rule, which defines the quantum metric $g$ \cite{Souza2000} and Chern number $\mathcal{C}$; the $\eta=1$ sum rule defines the plasma frequency or the effective optical mass $n/m_g$ (where $n$ is total density) and the orbital magnetic moment $\mu_M$ \cite{Souza2008}, and lastly $\eta=-1$ which defines $\chi_e$ and $\chi_m$. While $\chi_e$ is the electric susceptibility, related to the capacitance of the insulator \cite{komissarov2024quantum}, $\chi_m$ represents a torsion constant related to the chiral-magnetic effect.
As a consequence of Kramers-Kronig, the encircled quantities in Fig.~\ref{fig1:schematic}{\bf b} can be found in the frequency expansion of conductivity. If the insulating gap is $E_g$, the sub-gap conductivity admits the expression
\begin{equation}
    \sigma_{\mu\nu}(\omega \ll E_g) = \dfrac{e^2}{\hbar} \big( \mathcal{C} \epsilon_{\mu\nu} + \delta_{\mu\nu} i \omega \chi_e \big)
\end{equation}
featuring both $\mathcal{C}$ and $\chi_e$ \cite{komissarov2024quantum}. The optical mass also makes an appearance, albeit at high frequencies as an inductive piece $\sim i n/(m_g\omega)$.
Such high-frequency regimes are typically accessible only for superconductors, where $n/m_g$ is replaced by the superfluid stiffness \cite{Ferrell1958,Tinkham1959}.
In principle, $n/m_g$ appears in a nondissipative response (in this case at high frequency), in stark contrast to the quantum metric $g$, orbital magnetic moment $\mu_M$ and $\chi_m$, which can only be captured via sum rules over the entire spectrum.
As we emphasized earlier, the reason behind these omissions traces back to Eq.~\eqref{eq:sigma-Q-relation-t}, which includes only the anti-symmetric part of $\mathcal{Q}_{\mu\nu}(t)$.

\paragraph*{Step Response Theory.---}
Let us briefly recall the assumptions behind linear response theory.
We consider the system described by $\mathcal{H} = \mathcal{H}_0 - f(t) \hat{B}$ where $\mathcal{H}_0$ is the unperturbed Hamiltonian. The drive function $f(t)$ introduces a perturbation with operator $\hat{B}$. 
We assume that the drive is switched on slowly such that $f(t) \propto e^{-0^+ |t|}$ for $t \rightarrow -\infty$.
Assuming that the system is in its ground state at $t=-\infty$, linear response theory gives the expectation value of a different operator $\hat{A}$ as
\begin{equation}
    \langle \hat{A}(t) \rangle - \langle \hat{A}\rangle_0 = \int d t^\prime \chi_{AB}(t-t^\prime) f(t^\prime) \label{eq:A-t-linear-response}
\end{equation}
where the response function is given by the anti-symmetric part of the correlation function $\chi_{AB}(t-t^\prime) = i \Theta(t-t^\prime) \mathcal{R}^-_{AB}(t-t^\prime)$ where
\begin{equation}
    \mathcal{R}^-_{AB}(t) = \hat{A}(t) \hat{B}(0) - \hat{B}(0) \hat{A}(t) = \left\langle \left[ \hat{A}(t), \hat{B}(0) \right] \right\rangle.
\end{equation}
The $\Theta(t-t^\prime)$ enforces causality and the factor of $i$ makes $\chi_{AB}(t)$ hermitian.
We can further insert $f(t) = e^{ i (\omega_0 - i 0^+) t}$ to study the response function at a given driving frequency $\omega_0$.

The calculation of step response begins by considering the drive $f_{s}(t) = f_0 \Theta(-t)$ \cite{mazenko2006book}.
As depicted in Fig.~\ref{fig1:schematic}{\bf a}, the drive takes the system to a constrained equilibrium state from which it relaxes back to the unperturbed ground state as the drive is turned off.
We thus focus on $\hat{A}(t)$ for $t>0$, insert drive function $f_s(t)$ into Eq.~\eqref{eq:A-t-linear-response} and find 
\begin{equation}
    \langle \hat{A}(t) \rangle - \langle \hat{A} \rangle_0 = \mathcal{R}_{AB}(t) f_0
\end{equation}
where the relaxation function $\mathcal{R}_{AB}(t)$ is given by 
\begin{equation}
   \mathcal{R}_{AB}(t) = \int d\omega \;e^{-i \omega t} \; \dfrac{ \mathcal{R}^-_{AB}(\omega) }{ \omega - i 0^+ }. \label{eq:decay-d-t}
\end{equation}
This equation is perhaps not surprising since $\Theta(-t)$ is a highly singular function $\sim 1/\omega$ in the frequency domain. It thus mixes various frequency channels as a consequence of the convolution theorem.
Typically, $\mathcal{R}^-_{AB}/\omega$ is related to the Fourier transform of the anti-commutator $\mathcal{R}^+_{AB}(t) = \langle \{ \hat{A}(t), \hat{B}(0) \} \rangle$ by a fluctuation-dissipation relation \cite{mazenko2006book}. We derive these relations in the next section.

\paragraph*{Generalized fluctuation-dissipation relations.---}
We begin by recalling that the commutator $\mathcal{R}^-_{AB}(t)$ and the anti-commutator $\mathcal{R}^+_{AB}(t)$ can always be defined for a given correlation function $\mathcal{C}_{AB}(t) = \langle \hat{A}(t) \hat{B}(0) \rangle$. These quantities are not independent and are related by the fluctuation-dissipation theorem.
A standard derivation follows from the representation of these quantities on the exact basis
\begin{equation}
    \mathcal{R}^\pm_{AB}(t) = \sum\limits_{m,n} \mathcal{F}_{mn}^\pm e^{ i \hbar \omega_{mn} t } A_{nm} B_{mn} \label{eq:S-AB-t-full}
\end{equation}
where $\mathcal{F}_{mn}^\pm = f_m \pm f_n$ and $\{ f_n \}$ are the occupation factors.
Since these occupation factors follow $f_n/f_m =e^{-\beta \hbar\omega_{nm} }$, we can switch  to the Fourier domain by replacing $e^{ i \hbar \omega_{mn} t }$ with $\delta(\omega - \omega_{nm} )$ and derive the standard fluctuation-dissipation theorem \cite{Callen1951}
\begin{equation}
    \mathcal{R}^-_{AB}(\omega) = \tanh\left(\dfrac{\beta \hbar \omega}{2} \right) \mathcal{R}^+_{AB}(\omega). \label{eq:FDT-usual}
\end{equation}
The most important consequence of the fluctuation-dissipation relation is that the relaxation function in Eq.~\eqref{eq:decay-d-t} reduces to $\mathcal{R}^+_{AB}(t)$ in the classical limit where $\tanh(\beta \hbar\omega/2)\approx \beta\hbar\omega/2$, that is $\mathcal{R}_{AB}(t) = \beta \hbar \mathcal{R}_{AB}^+(t)$.
In sum, a combination of the fluctuation-dissipation relation and the classical limit provides a recipe to measure the anti-commutator $\mathcal{R}^+_{AB}(t)$.

We now explore whether the same formalism can be applied to measure the symmetric part of tQGT $\mathcal{Q}^s_{\mu\nu}(t)$.
We first notice that $\mathcal{Q}^{s}_{\mu\nu}(t)$ and $\mathcal{Q}^{as}_{\mu\nu}(t)$ have quite different pre-factors because of the complimentary projector. For instance, the symmetric part takes the expression
\begin{equation}
    \mathcal{Q}^s_{\mu\nu}(t) = \int_{\bf k} \sum\limits_{m\neq n} \mathcal{F}_{nm}^+ \hat{r}_\mu^{nm} \hat{r}_\nu^{mn} e^{ i \hbar \omega_{mn} t }
\end{equation}
where $\mathcal{F}_{nm}^\pm = f_n(1-f_m) \pm f_m(1-f_n)$. The anti-symmetric part similarly has
\begin{equation}
    \mathcal{F}_{nm}^- = f_n(1-f_m) - f_m(1-f_n) = f_n - f_m \label{eq:F-m-R-commutator}.
\end{equation}
These factors are different from those in Eq.~\eqref{eq:S-AB-t-full}.
However, owing to the fermionic nature of electrons, we can use the Fermi distribution functions that crucially satisfy
\begin{equation}
    \dfrac{ f_m(1-f_n) }{ f_n(1-f_m)  } = e^{ \beta \hbar \omega_{nm} }
\end{equation}
as a consequence of detailed balance condition for optical transitions. Therefore, despite complicated pre-factors, we get
\begin{equation}
    \mathcal{Q}^{as}_{\mu\nu}(\omega) = -\tanh\left(\dfrac{\beta \hbar \omega}{2} \right) \mathcal{Q}^s_{\mu\nu}(\omega). \label{eq:FDT-Q}
\end{equation}
The only difference between the usual fluctuation-dissipation theorem Eq.~\eqref{eq:FDT-usual} and the generalized one in Eq.~\eqref{eq:FDT-Q} is the minus sign.

\paragraph*{Step Electric Field.---}
We consider an external electric field $E_\nu$ which couples to the Hamiltonian via dipolar coupling $\mathcal{H} = \mathcal{H}_0 - e E_\nu(t) \hat{r}_\nu$
where $e$ is the electron charge and $E_\nu(t)$ describes the step drive shown in Fig.~\ref{fig1:schematic}{\bf a}. It is a constant $E_\nu$ for $t<0$ and zero for later times. We next compute the polarization from dipole moment $\mathcal{D}_\mu = -e \langle \hat{r}_\mu \rangle$ to linear order in the field and find
\begin{equation}
    \mathcal{D}_\mu(t) = -\left(e^2 \int d t^{\prime} i \Theta(t-t^\prime) \mathcal{R}^-_{\mu\nu}(t-t^\prime) f_s(t^\prime) \right) E_\nu
\end{equation}
where $\mathcal{R}^-_{\mu\nu}(t-t^\prime) = \langle [ \hat{r}_\mu(t), \hat{r}_\nu(t^\prime) ] \rangle $ is the usual commutator which can be written out explicitly in terms of Bloch states
\begin{equation}
    \mathcal{R}^-_{\mu\nu}(t) = \int_{\bf k} \sum\limits_{m\neq n} \mathcal{F}_{mn}^- \hat{r}_\mu^{nm} \hat{r}_\nu^{mn} e^{ i \hbar \omega_{mn} t }
\end{equation}
where $\mathcal{F}_{mn}^-$ are the Fermi factors defined in Eq.~\eqref{eq:F-m-R-commutator}.
We note that $\mathcal{R}^-_{\mu\nu}(t)$ is equal to the anti-symmetric part of tQGT $\mathcal{Q}^{as}_{\mu\nu}(t)$. As a result, the generalized fluctuation-dissipation relation in Eq.~\eqref{eq:FDT-Q} connects $\mathcal{Q}^{as}_{\mu\nu}(t)$ to the symmetric part $\mathcal{Q}^s_{\mu\nu}(t)$. Putting it all together, we find that the polarization $\mathcal{D}_\mu(t)$ responds to the step electric field $\mathcal{D}_\mu(t) = \mathcal{R}_{\mu\nu}(t) E_\nu$ where the relaxation function is given by
\begin{equation}
    \mathcal{R}_{\mu\nu}(t) =  \int d\omega \;e^{-i \omega t} \; \dfrac{ \tanh(\beta\hbar\omega/2) }{ \omega - i 0^+ } \mathcal{Q}^s_{\mu\nu}(\omega).\label{eq:main-result}
\end{equation}
This equation is the main result of our work. In the classical limit, $\beta \hbar \omega \ll 1$, where $\tanh(x)\sim x$ we get \begin{align}
    \mathcal{R}_{\mu\nu}(t) = \beta \hbar \mathcal{Q}^s_{\mu\nu}(t)/2.
\end{align}
In an experiment, the measurement will likely include the relaxation function from which the symmetric part of tQGT will need to be extracted. To that end, we must invert the integral using the convolution theorem. We find that $\mathcal{Q}^s_{\mu\nu}(t) = \int dt^\prime K(t-t^\prime)\;\mathcal{R}_{\mu\nu}(t^\prime)$ where the kernel is given by
\begin{equation}
    K(t) = \dfrac{ \sqrt{2\pi^3} }{ \beta^2\hbar^2} {\rm cosech}(\pi t/\beta\hbar)^2 .\label{eq:kernel-function}
\end{equation}

Lastly, we comment on different moments of the relaxation function. Similar to tQGT, different moments carry different information about different geometric properties of the system. The longitudinal component of the $t=0$ value is related to the quantum metric. Similarly, the Hall component of the derivative at $t=0$ is related to the orbital magnetic moment. In sum, the relaxation function thus acts as the generating function for the boxed objects in Fig.~\ref{fig1:schematic}{\bf b}.

\paragraph*{SSH chain.---}

\begin{figure}
    \centering
    \includegraphics[width=0.49\textwidth]{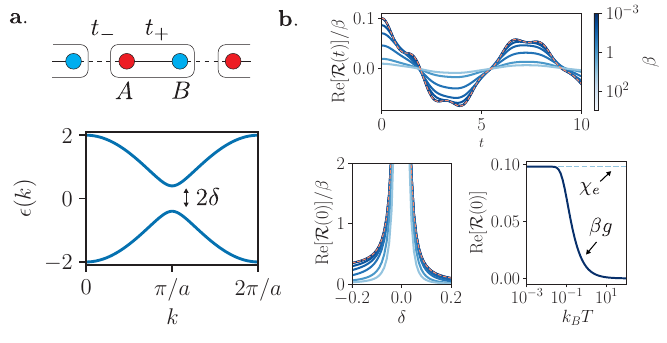}
    \caption{{\bf a}. 1D SSH chain with two orbitals $A$ and $B$ in the unit cell that is denoted by the box. The dashed lines indicate staggered hopping with $t_\pm = (1\pm \delta)$ which leads to a gap in the band structure. The system is an insulator at half-filling. {\bf b}. The relaxation function for different temperatures $\beta$ as a function of time. The orange dashed line is the symmetric part of the tQGT $\mathcal{Q}^s(t)$. As expected, the relaxation function $\mathcal{R}(t)$ coincides with $\beta \mathcal{Q}^s(t)$ in the classical limit $\beta \ll 1$. 
    It also follows the behavior as the system goes through the topological phase transition with $\delta$.
    Interestingly, $\mathcal{R}(t)$ approaches the susceptibility $\chi_e$ in the opposite quantum limit with $\beta \gg 1$.
    }
    \label{fig2_ssh}
\end{figure}

As a concrete example to illustrate the step response, we turn to the 1D SSH chain \cite{su1979}. We consider two orbitals in a unit cell described by the Hamiltonian
\begin{equation}
    \mathcal{H} = \sum\limits_{ i } (1+\delta) c^\dag_{i,A} c^\phdag_{i,B} + (1-\delta) c^\dag_{i,A} c^\phdag_{i-1,B} + {\rm h.c.}
\end{equation}
where $t$ is the hopping amplitude and $\delta$ denotes the imbalance between inter- and intra-cell hopping (see Fig.~\ref{fig2_ssh}{\bf a}), and $\hbar$ is set to 1. The Bloch Hamiltonian in momentum space can be written as $H_k = {\bf d}(k)\cdot \boldsymbol\sigma$ where $\{\sigma_i\}$ are the Pauli matrices and 
\begin{equation}
    d_x(k) = t_+ + t_- \cos k ,\; d_y(k) = t_- \sin k ,\; d_z(k) = 0
\end{equation}
where $t_\pm = (1\pm \delta)$. The staggering $\delta$ introduces a gap in the band structure, rendering the system an insulator at half-filling.

The insulator, however, has bounded oscillations, which manifest in a non-trivial tQGT. The symmetric part $\mathcal{Q}^s(t)$ is plotted in Fig.~\ref{fig2_ssh}{\bf b} with a dashed orange line. 
Importantly, we see that it can be closely approximated by the relaxation function in the classical limit.
The relaxation function in this model is given by
\begin{equation}
    \mathcal{R}(t) = \int dk\; \dfrac{ \tanh(\beta |{\bf d}(k)|) }{ 2|{\bf d}(k)| } \; g_{12}(k)  \;e^{ -i 2 |{\bf d}(k)| t }
\end{equation}
where $g_{12}(k) = |\langle u_{-,k} | \partial_k u_{+,k} \rangle|^2$ is the position matrix element between states that are defined by $H_k | u_{\pm, k}\rangle = \pm |{\bf d}(k)| |u_{\pm, k}\rangle$. The classical limit proceeds by replacing $\tanh(\beta |{\bf d}|) \rightarrow \beta |{\bf d}|$ which gives the intended result $\mathcal{R} = \beta \mathcal{Q}^s(t)/2$.

In the opposite limit, where $\beta \gg 1$ so that $\tanh(\beta |{\bf d}|) \rightarrow 1$, we find that $\mathcal{R}(0)$ approaches $\chi_e$
\begin{equation}
    \mathcal{R}(0) = \chi_e = \int dk\; \dfrac{g_{12}(k)}{ 2|{\bf d}(k)| }
\end{equation}
which has been recently related to the geometric capacitance in insulators \cite{komissarov2024quantum} since it enters in the steady state polarization $\mathcal{P} \propto \epsilon_0 \chi_e E$ and hence directly affects the energy stored $\sim \mathcal{P} E = \epsilon_0\chi_e E^2$ inside the material \cite{Kaplan2024}.

\paragraph*{Discussion.---}

The zero-point motion of bound electrons in insulators reflects quantum geometry and necessitates a nonvanishing quantum metric in the electron wavefunction.
However, probing the quantum metric directly in a material has remained a challenge. One of the reasons, as we outline in this work, is that linear response to a DC electric field will necessarily pick up the antisymmetric part of the dipole-dipole correlator for fermions, while the quantum metric belongs to the symmetric part.
We, therefore, propose a response function in which the electric field itself is antisymmetric in frequency.
In particular, we show that relaxation in the bulk dipole moment can directly probe the quantum metric. It can be understood as follows. The fully polarized the medium at $t=0$ exhibits longitudinal dipole-dipole oscillations as the field is turned off and the system relaxes back to its unperturbed equilibrium state.
The relaxation function thus obtained also captures the other geometric quantities that are hidden in linear response.
Heuristically, the step electric field $E(t)\propto\Theta(t)$ induces response at $1/\omega$ which is odd in frequency, and leverages the SWM sum rule to yield the quantum metric in the relaxation function $\mathcal{R}(t)$.

The relaxation function $\mathcal{R}(t)$ contains a convolution of the symmetric part of the tQGT $\mathcal{Q}^s(t)$ with $\tanh(\beta\omega)$ which reduces to $\mathcal{R}_{\mu\nu}(t) = \beta \hbar \mathcal{Q}^s_{\mu\nu}(t)/2$ in the classical high-temperature limit. The limit allows for the direct measurement of the quantum metric $g_{\mu\nu}=\Q_{\mu\nu}^s(t=0)$. In the opposite (low-temperature) limit, we found that the dipole relaxation function approaches the susceptibility $\chi_e$, which appears in nondissipative linear response. 
It remains to be seen whether such an experiment can be performed in a realistic setup. The high-temperature limit would be the most challenging to implement, as the temperature needs to be the largest energy scale in the system. There may be simplifications when a subset of bands form a closed subspace such that the high temperature limit requires temperatures higher than the bandwidth.
Lastly, we note that the kernel in Eq.\eqref{eq:kernel-function} is highly non-linear and error propagation can pose significant challenge to a realistic experiment.

\paragraph*{Acknowledgement.---}
Work on quantum geometric properties of quantum materials is supported as part of Programmable Quantum Materials, an Energy Frontier Research Center funded by the U.S. Department of Energy (DOE), Office of Science, Basic Energy Sciences (BES), under award DE-SC0019443. 
N.V.~greatly benefited from discussions with D. Kaplan during the Aspen Center for Physics workshop on ``Quantum Matter Through the Lens of Moiré Materials''.
The Flatiron Institute is a division of
the Simons Foundation.

\bibliography{qmetric}

\begin{thebibliography}{41}%
\makeatletter
\providecommand \@ifxundefined [1]{%
 \@ifx{#1\undefined}
}%
\providecommand \@ifnum [1]{%
 \ifnum #1\expandafter \@firstoftwo
 \else \expandafter \@secondoftwo
 \fi
}%
\providecommand \@ifx [1]{%
 \ifx #1\expandafter \@firstoftwo
 \else \expandafter \@secondoftwo
 \fi
}%
\providecommand \natexlab [1]{#1}%
\providecommand \enquote  [1]{``#1''}%
\providecommand \bibnamefont  [1]{#1}%
\providecommand \bibfnamefont [1]{#1}%
\providecommand \citenamefont [1]{#1}%
\providecommand \href@noop [0]{\@secondoftwo}%
\providecommand \href [0]{\begingroup \@sanitize@url \@href}%
\providecommand \@href[1]{\@@startlink{#1}\@@href}%
\providecommand \@@href[1]{\endgroup#1\@@endlink}%
\providecommand \@sanitize@url [0]{\catcode `\\12\catcode `\$12\catcode `\&12\catcode `\#12\catcode `\^12\catcode `\_12\catcode `\%12\relax}%
\providecommand \@@startlink[1]{}%
\providecommand \@@endlink[0]{}%
\providecommand \url  [0]{\begingroup\@sanitize@url \@url }%
\providecommand \@url [1]{\endgroup\@href {#1}{\urlprefix }}%
\providecommand \urlprefix  [0]{URL }%
\providecommand \Eprint [0]{\href }%
\providecommand \doibase [0]{https://doi.org/}%
\providecommand \selectlanguage [0]{\@gobble}%
\providecommand \bibinfo  [0]{\@secondoftwo}%
\providecommand \bibfield  [0]{\@secondoftwo}%
\providecommand \translation [1]{[#1]}%
\providecommand \BibitemOpen [0]{}%
\providecommand \bibitemStop [0]{}%
\providecommand \bibitemNoStop [0]{.\EOS\space}%
\providecommand \EOS [0]{\spacefactor3000\relax}%
\providecommand \BibitemShut  [1]{\csname bibitem#1\endcsname}%
\let\auto@bib@innerbib\@empty
\bibitem [{\citenamefont {Hasan}\ and\ \citenamefont {Kane}(2010)}]{Hasan2010}%
  \BibitemOpen
  \bibfield  {author} {\bibinfo {author} {\bibfnamefont {M.~Z.}\ \bibnamefont {Hasan}}\ and\ \bibinfo {author} {\bibfnamefont {C.~L.}\ \bibnamefont {Kane}},\ }\href {https://doi.org/10.1103/RevModPhys.82.3045} {\bibfield  {journal} {\bibinfo  {journal} {Rev. Mod. Phys.}\ }\textbf {\bibinfo {volume} {82}},\ \bibinfo {pages} {3045} (\bibinfo {year} {2010})}\BibitemShut {NoStop}%
\bibitem [{\citenamefont {Monaco}\ \emph {et~al.}(2018)\citenamefont {Monaco}, \citenamefont {Panati}, \citenamefont {Pisante},\ and\ \citenamefont {Teufel}}]{Monaco2018}%
  \BibitemOpen
  \bibfield  {author} {\bibinfo {author} {\bibfnamefont {D.}~\bibnamefont {Monaco}}, \bibinfo {author} {\bibfnamefont {G.}~\bibnamefont {Panati}}, \bibinfo {author} {\bibfnamefont {A.}~\bibnamefont {Pisante}},\ and\ \bibinfo {author} {\bibfnamefont {S.}~\bibnamefont {Teufel}},\ }\href {https://doi.org/10.1007/s00220-017-3067-7} {\bibfield  {journal} {\bibinfo  {journal} {Communications in Mathematical Physics}\ }\textbf {\bibinfo {volume} {359}},\ \bibinfo {pages} {61} (\bibinfo {year} {2018})}\BibitemShut {NoStop}%
\bibitem [{\citenamefont {Verma}\ and\ \citenamefont {Queiroz}(2024)}]{verma2024instantaneous}%
  \BibitemOpen
  \bibfield  {author} {\bibinfo {author} {\bibfnamefont {N.}~\bibnamefont {Verma}}\ and\ \bibinfo {author} {\bibfnamefont {R.}~\bibnamefont {Queiroz}},\ }\href {https://arxiv.org/abs/2403.07052} {\bibfield  {journal} {\bibinfo  {journal} {arXiv:2403.07052}\ } (\bibinfo {year} {2024})}\BibitemShut {NoStop}%
\bibitem [{\citenamefont {Onishi}\ and\ \citenamefont {Fu}(2024{\natexlab{a}})}]{onishi2024quantum}%
  \BibitemOpen
  \bibfield  {author} {\bibinfo {author} {\bibfnamefont {Y.}~\bibnamefont {Onishi}}\ and\ \bibinfo {author} {\bibfnamefont {L.}~\bibnamefont {Fu}},\ }\href {https://arxiv.org/abs/2401.13847} {\bibfield  {journal} {\bibinfo  {journal} {arXiv preprint arXiv:2401.13847}\ } (\bibinfo {year} {2024}{\natexlab{a}})}\BibitemShut {NoStop}%
\bibitem [{\citenamefont {Julku}\ \emph {et~al.}(2016)\citenamefont {Julku}, \citenamefont {Peotta}, \citenamefont {Vanhala}, \citenamefont {Kim},\ and\ \citenamefont {T\"orm\"a}}]{Julku2016}%
  \BibitemOpen
  \bibfield  {author} {\bibinfo {author} {\bibfnamefont {A.}~\bibnamefont {Julku}}, \bibinfo {author} {\bibfnamefont {S.}~\bibnamefont {Peotta}}, \bibinfo {author} {\bibfnamefont {T.~I.}\ \bibnamefont {Vanhala}}, \bibinfo {author} {\bibfnamefont {D.-H.}\ \bibnamefont {Kim}},\ and\ \bibinfo {author} {\bibfnamefont {P.}~\bibnamefont {T\"orm\"a}},\ }\href {https://doi.org/10.1103/PhysRevLett.117.045303} {\bibfield  {journal} {\bibinfo  {journal} {Phys. Rev. Lett.}\ }\textbf {\bibinfo {volume} {117}},\ \bibinfo {pages} {045303} (\bibinfo {year} {2016})}\BibitemShut {NoStop}%
\bibitem [{\citenamefont {Törmä}\ \emph {et~al.}(2022)\citenamefont {Törmä}, \citenamefont {Peotta},\ and\ \citenamefont {Bernevig}}]{Torma2021}%
  \BibitemOpen
  \bibfield  {author} {\bibinfo {author} {\bibfnamefont {P.}~\bibnamefont {Törmä}}, \bibinfo {author} {\bibfnamefont {S.}~\bibnamefont {Peotta}},\ and\ \bibinfo {author} {\bibfnamefont {B.~A.}\ \bibnamefont {Bernevig}},\ }\href {https://doi.org/10.1038/s42254-022-00466-y} {\bibfield  {journal} {\bibinfo  {journal} {Nat. Rev. Phys.}\ }\textbf {\bibinfo {volume} {4}},\ \bibinfo {pages} {528} (\bibinfo {year} {2022})}\BibitemShut {NoStop}%
\bibitem [{\citenamefont {Yu}\ \emph {et~al.}(2023)\citenamefont {Yu}, \citenamefont {Ciccarino}, \citenamefont {Bianco}, \citenamefont {Errea}, \citenamefont {Narang},\ and\ \citenamefont {Bernevig}}]{yu2023}%
  \BibitemOpen
  \bibfield  {author} {\bibinfo {author} {\bibfnamefont {J.}~\bibnamefont {Yu}}, \bibinfo {author} {\bibfnamefont {C.~J.}\ \bibnamefont {Ciccarino}}, \bibinfo {author} {\bibfnamefont {R.}~\bibnamefont {Bianco}}, \bibinfo {author} {\bibfnamefont {I.}~\bibnamefont {Errea}}, \bibinfo {author} {\bibfnamefont {P.}~\bibnamefont {Narang}},\ and\ \bibinfo {author} {\bibfnamefont {B.~A.}\ \bibnamefont {Bernevig}},\ }\href {https://arxiv.org/abs/2305.02340} {\bibfield  {journal} {\bibinfo  {journal} {arXiv}\ ,\ \bibinfo {pages} {2305.02340}} (\bibinfo {year} {2023})}\BibitemShut {NoStop}%
\bibitem [{\citenamefont {Aoki}(2020)}]{Aoki2020}%
  \BibitemOpen
  \bibfield  {author} {\bibinfo {author} {\bibfnamefont {H.}~\bibnamefont {Aoki}},\ }\href {https://doi.org/10.1007/s10948-020-05474-6} {\bibfield  {journal} {\bibinfo  {journal} {Journal of Superconductivity and Novel Magnetism}\ }\textbf {\bibinfo {volume} {33}},\ \bibinfo {pages} {2341} (\bibinfo {year} {2020})}\BibitemShut {NoStop}%
\bibitem [{\citenamefont {Andrei}\ \emph {et~al.}(2021)\citenamefont {Andrei}, \citenamefont {Efetov}, \citenamefont {Jarillo-Herrero}, \citenamefont {MacDonald}, \citenamefont {Mak}, \citenamefont {Senthil}, \citenamefont {Tutuc}, \citenamefont {Yazdani},\ and\ \citenamefont {Young}}]{Andrei2021}%
  \BibitemOpen
  \bibfield  {author} {\bibinfo {author} {\bibfnamefont {E.~Y.}\ \bibnamefont {Andrei}}, \bibinfo {author} {\bibfnamefont {D.~K.}\ \bibnamefont {Efetov}}, \bibinfo {author} {\bibfnamefont {P.}~\bibnamefont {Jarillo-Herrero}}, \bibinfo {author} {\bibfnamefont {A.~H.}\ \bibnamefont {MacDonald}}, \bibinfo {author} {\bibfnamefont {K.~F.}\ \bibnamefont {Mak}}, \bibinfo {author} {\bibfnamefont {T.}~\bibnamefont {Senthil}}, \bibinfo {author} {\bibfnamefont {E.}~\bibnamefont {Tutuc}}, \bibinfo {author} {\bibfnamefont {A.}~\bibnamefont {Yazdani}},\ and\ \bibinfo {author} {\bibfnamefont {A.~F.}\ \bibnamefont {Young}},\ }\href {https://doi.org/10.1038/s41578-021-00284-1} {\bibfield  {journal} {\bibinfo  {journal} {Nat. Rev. Mat.}\ }\textbf {\bibinfo {volume} {6}},\ \bibinfo {pages} {201} (\bibinfo {year} {2021})}\BibitemShut {NoStop}%
\bibitem [{\citenamefont {Abouelkomsan}\ \emph {et~al.}(2023)\citenamefont {Abouelkomsan}, \citenamefont {Yang},\ and\ \citenamefont {Bergholtz}}]{Abouelkomsan2023}%
  \BibitemOpen
  \bibfield  {author} {\bibinfo {author} {\bibfnamefont {A.}~\bibnamefont {Abouelkomsan}}, \bibinfo {author} {\bibfnamefont {K.}~\bibnamefont {Yang}},\ and\ \bibinfo {author} {\bibfnamefont {E.~J.}\ \bibnamefont {Bergholtz}},\ }\href {https://doi.org/10.1103/PhysRevResearch.5.L012015} {\bibfield  {journal} {\bibinfo  {journal} {Phys. Rev. Res.}\ }\textbf {\bibinfo {volume} {5}},\ \bibinfo {pages} {L012015} (\bibinfo {year} {2023})}\BibitemShut {NoStop}%
\bibitem [{\citenamefont {Kwan}\ \emph {et~al.}(2021)\citenamefont {Kwan}, \citenamefont {Hu}, \citenamefont {Simon},\ and\ \citenamefont {Parameswaran}}]{kwan2021}%
  \BibitemOpen
  \bibfield  {author} {\bibinfo {author} {\bibfnamefont {Y.~H.}\ \bibnamefont {Kwan}}, \bibinfo {author} {\bibfnamefont {Y.}~\bibnamefont {Hu}}, \bibinfo {author} {\bibfnamefont {S.~H.}\ \bibnamefont {Simon}},\ and\ \bibinfo {author} {\bibfnamefont {S.~A.}\ \bibnamefont {Parameswaran}},\ }\href {https://doi.org/10.1103/PhysRevLett.126.137601} {\bibfield  {journal} {\bibinfo  {journal} {Phys. Rev. Lett.}\ }\textbf {\bibinfo {volume} {126}},\ \bibinfo {pages} {137601} (\bibinfo {year} {2021})}\BibitemShut {NoStop}%
\bibitem [{\citenamefont {Souza}\ \emph {et~al.}(2000)\citenamefont {Souza}, \citenamefont {Wilkens},\ and\ \citenamefont {Martin}}]{Souza2000}%
  \BibitemOpen
  \bibfield  {author} {\bibinfo {author} {\bibfnamefont {I.}~\bibnamefont {Souza}}, \bibinfo {author} {\bibfnamefont {T.}~\bibnamefont {Wilkens}},\ and\ \bibinfo {author} {\bibfnamefont {R.~M.}\ \bibnamefont {Martin}},\ }\href {https://doi.org/10.1103/PhysRevB.62.1666} {\bibfield  {journal} {\bibinfo  {journal} {Phys. Rev. B}\ }\textbf {\bibinfo {volume} {62}},\ \bibinfo {pages} {1666} (\bibinfo {year} {2000})}\BibitemShut {NoStop}%
\bibitem [{\citenamefont {Thonhauser}\ \emph {et~al.}(2005)\citenamefont {Thonhauser}, \citenamefont {Ceresoli}, \citenamefont {Vanderbilt},\ and\ \citenamefont {Resta}}]{Thonhauser2005}%
  \BibitemOpen
  \bibfield  {author} {\bibinfo {author} {\bibfnamefont {T.}~\bibnamefont {Thonhauser}}, \bibinfo {author} {\bibfnamefont {D.}~\bibnamefont {Ceresoli}}, \bibinfo {author} {\bibfnamefont {D.}~\bibnamefont {Vanderbilt}},\ and\ \bibinfo {author} {\bibfnamefont {R.}~\bibnamefont {Resta}},\ }\href {https://doi.org/10.1103/PhysRevLett.95.137205} {\bibfield  {journal} {\bibinfo  {journal} {Phys. Rev. Lett.}\ }\textbf {\bibinfo {volume} {95}},\ \bibinfo {pages} {137205} (\bibinfo {year} {2005})}\BibitemShut {NoStop}%
\bibitem [{\citenamefont {Souza}\ and\ \citenamefont {Vanderbilt}(2008)}]{Souza2008}%
  \BibitemOpen
  \bibfield  {author} {\bibinfo {author} {\bibfnamefont {I.}~\bibnamefont {Souza}}\ and\ \bibinfo {author} {\bibfnamefont {D.}~\bibnamefont {Vanderbilt}},\ }\href {https://doi.org/10.1103/PhysRevB.77.054438} {\bibfield  {journal} {\bibinfo  {journal} {Phys. Rev. B}\ }\textbf {\bibinfo {volume} {77}},\ \bibinfo {pages} {054438} (\bibinfo {year} {2008})}\BibitemShut {NoStop}%
\bibitem [{\citenamefont {Carmichael}(1999)}]{carmichael1999two}%
  \BibitemOpen
  \bibfield  {author} {\bibinfo {author} {\bibfnamefont {H.~J.}\ \bibnamefont {Carmichael}},\ }\href {https://link.springer.com/chapter/10.1007/978-3-662-03875-8_2} {\emph {\bibinfo {title} {Two-Level Atoms and Spontaneous Emission}}}\ (\bibinfo  {publisher} {Springer},\ \bibinfo {year} {1999})\ pp.\ \bibinfo {pages} {29--74}\BibitemShut {NoStop}%
\bibitem [{\citenamefont {Komissarov}\ \emph {et~al.}(2024)\citenamefont {Komissarov}, \citenamefont {Holder},\ and\ \citenamefont {Queiroz}}]{komissarov2024quantum}%
  \BibitemOpen
  \bibfield  {author} {\bibinfo {author} {\bibfnamefont {I.}~\bibnamefont {Komissarov}}, \bibinfo {author} {\bibfnamefont {T.}~\bibnamefont {Holder}},\ and\ \bibinfo {author} {\bibfnamefont {R.}~\bibnamefont {Queiroz}},\ }\href {https://www.nature.com/articles/s41467-024-48808-x} {\bibfield  {journal} {\bibinfo  {journal} {Nat. Commun.}\ }\textbf {\bibinfo {volume} {15}},\ \bibinfo {pages} {4621} (\bibinfo {year} {2024})}\BibitemShut {NoStop}%
\bibitem [{\citenamefont {Verma}\ \emph {et~al.}(2021)\citenamefont {Verma}, \citenamefont {Hazra},\ and\ \citenamefont {Randeria}}]{Verma2021}%
  \BibitemOpen
  \bibfield  {author} {\bibinfo {author} {\bibfnamefont {N.}~\bibnamefont {Verma}}, \bibinfo {author} {\bibfnamefont {T.}~\bibnamefont {Hazra}},\ and\ \bibinfo {author} {\bibfnamefont {M.}~\bibnamefont {Randeria}},\ }\href {https://www.pnas.org/doi/10.1073/pnas.2106744118} {\bibfield  {journal} {\bibinfo  {journal} {Proc. Nat. Acad. Sci.}\ }\textbf {\bibinfo {volume} {118}} (\bibinfo {year} {2021})}\BibitemShut {NoStop}%
\bibitem [{\citenamefont {Mao}\ and\ \citenamefont {Chowdhury}(2023)}]{Mao2023}%
  \BibitemOpen
  \bibfield  {author} {\bibinfo {author} {\bibfnamefont {D.}~\bibnamefont {Mao}}\ and\ \bibinfo {author} {\bibfnamefont {D.}~\bibnamefont {Chowdhury}},\ }\href {https://doi.org/10.1073/pnas.2217816120} {\bibfield  {journal} {\bibinfo  {journal} {Proc. Natl. Acad. Sci. U.S.A.}\ }\textbf {\bibinfo {volume} {120}},\ \bibinfo {pages} {e2217816120} (\bibinfo {year} {2023})}\BibitemShut {NoStop}%
\bibitem [{\citenamefont {Kruchkov}\ and\ \citenamefont {Ryu}(2023)}]{Kruchkov2023}%
  \BibitemOpen
  \bibfield  {author} {\bibinfo {author} {\bibfnamefont {A.}~\bibnamefont {Kruchkov}}\ and\ \bibinfo {author} {\bibfnamefont {S.}~\bibnamefont {Ryu}},\ }\href {https://arxiv.org/abs/2312.17318} {\bibfield  {journal} {\bibinfo  {journal} {arXiv preprint arXiv:2312.17318}\ } (\bibinfo {year} {2023})}\BibitemShut {NoStop}%
\bibitem [{\citenamefont {Onishi}\ and\ \citenamefont {Fu}(2024{\natexlab{b}})}]{onishi2023quantum}%
  \BibitemOpen
  \bibfield  {author} {\bibinfo {author} {\bibfnamefont {Y.}~\bibnamefont {Onishi}}\ and\ \bibinfo {author} {\bibfnamefont {L.}~\bibnamefont {Fu}},\ }\href {https://doi.org/10.1103/PhysRevX.14.011052} {\bibfield  {journal} {\bibinfo  {journal} {Phys. Rev. X}\ }\textbf {\bibinfo {volume} {14}},\ \bibinfo {pages} {011052} (\bibinfo {year} {2024}{\natexlab{b}})}\BibitemShut {NoStop}%
\bibitem [{\citenamefont {Onishi}\ and\ \citenamefont {Fu}(2024{\natexlab{c}})}]{onishi2024universal}%
  \BibitemOpen
  \bibfield  {author} {\bibinfo {author} {\bibfnamefont {Y.}~\bibnamefont {Onishi}}\ and\ \bibinfo {author} {\bibfnamefont {L.}~\bibnamefont {Fu}},\ }\href {https://arxiv.org/abs/2401.04180} {\bibfield  {journal} {\bibinfo  {journal} {arXiv preprint arXiv:2401.04180}\ } (\bibinfo {year} {2024}{\natexlab{c}})}\BibitemShut {NoStop}%
\bibitem [{\citenamefont {Ghosh}\ \emph {et~al.}(2024)\citenamefont {Ghosh}, \citenamefont {Onishi}, \citenamefont {Xu}, \citenamefont {Lin}, \citenamefont {Fu},\ and\ \citenamefont {Bansil}}]{ghosh2024probing}%
  \BibitemOpen
  \bibfield  {author} {\bibinfo {author} {\bibfnamefont {B.}~\bibnamefont {Ghosh}}, \bibinfo {author} {\bibfnamefont {Y.}~\bibnamefont {Onishi}}, \bibinfo {author} {\bibfnamefont {S.-Y.}\ \bibnamefont {Xu}}, \bibinfo {author} {\bibfnamefont {H.}~\bibnamefont {Lin}}, \bibinfo {author} {\bibfnamefont {L.}~\bibnamefont {Fu}},\ and\ \bibinfo {author} {\bibfnamefont {A.}~\bibnamefont {Bansil}},\ }\href {https://arxiv.org/abs/2401.09689} {\bibfield  {journal} {\bibinfo  {journal} {arXiv preprint arXiv:2401.09689}\ } (\bibinfo {year} {2024})}\BibitemShut {NoStop}%
\bibitem [{\citenamefont {Ahn}\ \emph {et~al.}(2022)\citenamefont {Ahn}, \citenamefont {Guo}, \citenamefont {Nagaosa},\ and\ \citenamefont {Vishwanath}}]{Ahn2022}%
  \BibitemOpen
  \bibfield  {author} {\bibinfo {author} {\bibfnamefont {J.}~\bibnamefont {Ahn}}, \bibinfo {author} {\bibfnamefont {G.-Y.}\ \bibnamefont {Guo}}, \bibinfo {author} {\bibfnamefont {N.}~\bibnamefont {Nagaosa}},\ and\ \bibinfo {author} {\bibfnamefont {A.}~\bibnamefont {Vishwanath}},\ }\href {https://doi.org/10.1038/s41567-021-01465-z} {\bibfield  {journal} {\bibinfo  {journal} {Nat. Phys.}\ }\textbf {\bibinfo {volume} {18}},\ \bibinfo {pages} {290} (\bibinfo {year} {2022})}\BibitemShut {NoStop}%
\bibitem [{\citenamefont {Hofmann}\ \emph {et~al.}(2020)\citenamefont {Hofmann}, \citenamefont {Berg},\ and\ \citenamefont {Chowdhury}}]{Hofmann2020}%
  \BibitemOpen
  \bibfield  {author} {\bibinfo {author} {\bibfnamefont {J.~S.}\ \bibnamefont {Hofmann}}, \bibinfo {author} {\bibfnamefont {E.}~\bibnamefont {Berg}},\ and\ \bibinfo {author} {\bibfnamefont {D.}~\bibnamefont {Chowdhury}},\ }\href {https://doi.org/10.1103/PhysRevB.102.201112} {\bibfield  {journal} {\bibinfo  {journal} {Phys. Rev. B}\ }\textbf {\bibinfo {volume} {102}},\ \bibinfo {pages} {201112} (\bibinfo {year} {2020})}\BibitemShut {NoStop}%
\bibitem [{\citenamefont {Törmä}\ \emph {et~al.}(2018)\citenamefont {Törmä}, \citenamefont {Liang},\ and\ \citenamefont {Peotta}}]{Torma2018}%
  \BibitemOpen
  \bibfield  {author} {\bibinfo {author} {\bibfnamefont {P.}~\bibnamefont {Törmä}}, \bibinfo {author} {\bibfnamefont {L.}~\bibnamefont {Liang}},\ and\ \bibinfo {author} {\bibfnamefont {S.}~\bibnamefont {Peotta}},\ }\href {https://doi.org/10.1103/PhysRevB.98.220511} {\bibfield  {journal} {\bibinfo  {journal} {Phys. Rev. B}\ }\textbf {\bibinfo {volume} {98}},\ \bibinfo {pages} {220511} (\bibinfo {year} {2018})}\BibitemShut {NoStop}%
\bibitem [{\citenamefont {Iskin}(2018)}]{Iskin2018}%
  \BibitemOpen
  \bibfield  {author} {\bibinfo {author} {\bibfnamefont {M.}~\bibnamefont {Iskin}},\ }\href {https://doi.org/10.1103/PhysRevA.97.033625} {\bibfield  {journal} {\bibinfo  {journal} {Phys. Rev. A}\ }\textbf {\bibinfo {volume} {97}},\ \bibinfo {pages} {033625} (\bibinfo {year} {2018})}\BibitemShut {NoStop}%
\bibitem [{\citenamefont {Iskin}(2021)}]{Iskin2021}%
  \BibitemOpen
  \bibfield  {author} {\bibinfo {author} {\bibfnamefont {M.}~\bibnamefont {Iskin}},\ }\href {https://doi.org/10.1103/PhysRevA.103.053311} {\bibfield  {journal} {\bibinfo  {journal} {Phys. Rev. A}\ }\textbf {\bibinfo {volume} {103}},\ \bibinfo {pages} {053311} (\bibinfo {year} {2021})}\BibitemShut {NoStop}%
\bibitem [{\citenamefont {Ferrell}\ and\ \citenamefont {Glover}(1958)}]{Ferrell1958}%
  \BibitemOpen
  \bibfield  {author} {\bibinfo {author} {\bibfnamefont {R.~A.}\ \bibnamefont {Ferrell}}\ and\ \bibinfo {author} {\bibfnamefont {R.~E.}\ \bibnamefont {Glover}},\ }\href {https://doi.org/10.1103/PhysRev.109.1398} {\bibfield  {journal} {\bibinfo  {journal} {Phys. Rev.}\ }\textbf {\bibinfo {volume} {109}},\ \bibinfo {pages} {1398} (\bibinfo {year} {1958})}\BibitemShut {NoStop}%
\bibitem [{\citenamefont {Tinkham}\ and\ \citenamefont {Ferrell}(1959)}]{Tinkham1959}%
  \BibitemOpen
  \bibfield  {author} {\bibinfo {author} {\bibfnamefont {M.}~\bibnamefont {Tinkham}}\ and\ \bibinfo {author} {\bibfnamefont {R.~A.}\ \bibnamefont {Ferrell}},\ }\href {https://doi.org/10.1103/PhysRevLett.2.331} {\bibfield  {journal} {\bibinfo  {journal} {Phys. Rev. Lett.}\ }\textbf {\bibinfo {volume} {2}},\ \bibinfo {pages} {331} (\bibinfo {year} {1959})}\BibitemShut {NoStop}%
\bibitem [{\citenamefont {Hu}\ \emph {et~al.}(2022)\citenamefont {Hu}, \citenamefont {Hyart}, \citenamefont {Pikulin},\ and\ \citenamefont {Rossi}}]{Hu2022}%
  \BibitemOpen
  \bibfield  {author} {\bibinfo {author} {\bibfnamefont {X.}~\bibnamefont {Hu}}, \bibinfo {author} {\bibfnamefont {T.}~\bibnamefont {Hyart}}, \bibinfo {author} {\bibfnamefont {D.~I.}\ \bibnamefont {Pikulin}},\ and\ \bibinfo {author} {\bibfnamefont {E.}~\bibnamefont {Rossi}},\ }\href {https://doi.org/10.1103/PhysRevB.105.L140506} {\bibfield  {journal} {\bibinfo  {journal} {Phys. Rev. B}\ }\textbf {\bibinfo {volume} {105}},\ \bibinfo {pages} {L140506} (\bibinfo {year} {2022})}\BibitemShut {NoStop}%
\bibitem [{\citenamefont {Hu}\ \emph {et~al.}(2023)\citenamefont {Hu}, \citenamefont {Rossi},\ and\ \citenamefont {Barlas}}]{Hu2023}%
  \BibitemOpen
  \bibfield  {author} {\bibinfo {author} {\bibfnamefont {X.}~\bibnamefont {Hu}}, \bibinfo {author} {\bibfnamefont {E.}~\bibnamefont {Rossi}},\ and\ \bibinfo {author} {\bibfnamefont {Y.}~\bibnamefont {Barlas}},\ }\href {https://arxiv.org/abs/2304.04825} {\bibfield  {journal} {\bibinfo  {journal} {arXiv preprint arXiv:2304.04825}\ } (\bibinfo {year} {2023})}\BibitemShut {NoStop}%
\bibitem [{\citenamefont {Verma}\ \emph {et~al.}(2024)\citenamefont {Verma}, \citenamefont {Guerci},\ and\ \citenamefont {Queiroz}}]{Verma2024exciton}%
  \BibitemOpen
  \bibfield  {author} {\bibinfo {author} {\bibfnamefont {N.}~\bibnamefont {Verma}}, \bibinfo {author} {\bibfnamefont {D.}~\bibnamefont {Guerci}},\ and\ \bibinfo {author} {\bibfnamefont {R.}~\bibnamefont {Queiroz}},\ }\href {https://doi.org/10.1103/PhysRevLett.132.236001} {\bibfield  {journal} {\bibinfo  {journal} {Phys. Rev. Lett.}\ }\textbf {\bibinfo {volume} {132}},\ \bibinfo {pages} {236001} (\bibinfo {year} {2024})}\BibitemShut {NoStop}%
\bibitem [{\citenamefont {Mitscherling}\ and\ \citenamefont {Holder}(2022)}]{Mitscherling2022}%
  \BibitemOpen
  \bibfield  {author} {\bibinfo {author} {\bibfnamefont {J.}~\bibnamefont {Mitscherling}}\ and\ \bibinfo {author} {\bibfnamefont {T.}~\bibnamefont {Holder}},\ }\href {https://doi.org/10.1103/PhysRevB.105.085154} {\bibfield  {journal} {\bibinfo  {journal} {Phys. Rev. B}\ }\textbf {\bibinfo {volume} {105}},\ \bibinfo {pages} {085154} (\bibinfo {year} {2022})}\BibitemShut {NoStop}%
\bibitem [{\citenamefont {Huhtinen}\ and\ \citenamefont {T\"orm\"a}(2023)}]{Huhtinen2023}%
  \BibitemOpen
  \bibfield  {author} {\bibinfo {author} {\bibfnamefont {K.-E.}\ \bibnamefont {Huhtinen}}\ and\ \bibinfo {author} {\bibfnamefont {P.}~\bibnamefont {T\"orm\"a}},\ }\href {https://doi.org/10.1103/PhysRevB.108.155108} {\bibfield  {journal} {\bibinfo  {journal} {Phys. Rev. B}\ }\textbf {\bibinfo {volume} {108}},\ \bibinfo {pages} {155108} (\bibinfo {year} {2023})}\BibitemShut {NoStop}%
\bibitem [{\citenamefont {Resta}(2006)}]{Resta2006}%
  \BibitemOpen
  \bibfield  {author} {\bibinfo {author} {\bibfnamefont {R.}~\bibnamefont {Resta}},\ }\href {https://doi.org/10.1103/PhysRevLett.96.137601} {\bibfield  {journal} {\bibinfo  {journal} {Phys. Rev. Lett.}\ }\textbf {\bibinfo {volume} {96}},\ \bibinfo {pages} {137601} (\bibinfo {year} {2006})}\BibitemShut {NoStop}%
\bibitem [{\citenamefont {Resta}(2011)}]{Resta2011}%
  \BibitemOpen
  \bibfield  {author} {\bibinfo {author} {\bibfnamefont {R.}~\bibnamefont {Resta}},\ }\href {https://link.springer.com/article/10.1140/epjb/e2010-10874-4} {\bibfield  {journal} {\bibinfo  {journal} {The European Physical Journal B}\ }\textbf {\bibinfo {volume} {79}},\ \bibinfo {pages} {121} (\bibinfo {year} {2011})}\BibitemShut {NoStop}%
\bibitem [{\citenamefont {Blount}(1962)}]{Blount1962}%
  \BibitemOpen
  \bibfield  {author} {\bibinfo {author} {\bibfnamefont {E.~I.}\ \bibnamefont {Blount}},\ }\href {https://doi.org/10.1103/PhysRev.126.1636} {\bibfield  {journal} {\bibinfo  {journal} {Phys. Rev.}\ }\textbf {\bibinfo {volume} {126}},\ \bibinfo {pages} {1636} (\bibinfo {year} {1962})}\BibitemShut {NoStop}%
\bibitem [{\citenamefont {Mazenko}(2006)}]{mazenko2006book}%
  \BibitemOpen
  \bibfield  {author} {\bibinfo {author} {\bibfnamefont {G.}~\bibnamefont {Mazenko}},\ }\href {http://doi.org/10.1002/9783527618958} {\bibfield  {journal} {\bibinfo  {journal} {{Nonequilibrium Statistical Mechanics} (John Wiley \& Sons)}\ } (\bibinfo {year} {2006})}\BibitemShut {NoStop}%
\bibitem [{\citenamefont {Callen}\ and\ \citenamefont {Welton}(1951)}]{Callen1951}%
  \BibitemOpen
  \bibfield  {author} {\bibinfo {author} {\bibfnamefont {H.~B.}\ \bibnamefont {Callen}}\ and\ \bibinfo {author} {\bibfnamefont {T.~A.}\ \bibnamefont {Welton}},\ }\href {https://doi.org/10.1103/PhysRev.83.34} {\bibfield  {journal} {\bibinfo  {journal} {Phys. Rev.}\ }\textbf {\bibinfo {volume} {83}},\ \bibinfo {pages} {34} (\bibinfo {year} {1951})}\BibitemShut {NoStop}%
\bibitem [{\citenamefont {Su}\ \emph {et~al.}(1979)\citenamefont {Su}, \citenamefont {Schrieffer},\ and\ \citenamefont {Heeger}}]{su1979}%
  \BibitemOpen
  \bibfield  {author} {\bibinfo {author} {\bibfnamefont {W.~P.}\ \bibnamefont {Su}}, \bibinfo {author} {\bibfnamefont {J.~R.}\ \bibnamefont {Schrieffer}},\ and\ \bibinfo {author} {\bibfnamefont {A.~J.}\ \bibnamefont {Heeger}},\ }\href {https://doi.org/10.1103/PhysRevLett.42.1698} {\bibfield  {journal} {\bibinfo  {journal} {Phys. Rev. Lett.}\ }\textbf {\bibinfo {volume} {42}},\ \bibinfo {pages} {1698} (\bibinfo {year} {1979})}\BibitemShut {NoStop}%
\bibitem [{\citenamefont {Kaplan}\ \emph {et~al.}(2024)\citenamefont {Kaplan}, \citenamefont {Holder},\ and\ \citenamefont {Yan}}]{Kaplan2024}%
  \BibitemOpen
  \bibfield  {author} {\bibinfo {author} {\bibfnamefont {D.}~\bibnamefont {Kaplan}}, \bibinfo {author} {\bibfnamefont {T.}~\bibnamefont {Holder}},\ and\ \bibinfo {author} {\bibfnamefont {B.}~\bibnamefont {Yan}},\ }\href {https://doi.org/10.1103/PhysRevLett.132.026301} {\bibfield  {journal} {\bibinfo  {journal} {Phys. Rev. Lett.}\ }\textbf {\bibinfo {volume} {132}},\ \bibinfo {pages} {026301} (\bibinfo {year} {2024})}\BibitemShut {NoStop}%
\end{thebibliography}%

\end{document}